\documentclass[12pt]{article}

\usepackage{amssymb,amsmath}
\usepackage{fullpage}
\usepackage{graphicx}
\usepackage{amsmath}		
\usepackage[margin=1.0in]{geometry}
\usepackage{setspace}
\usepackage{color}
\usepackage{fancyhdr}
\usepackage{collcell}
\usepackage{datatool}
\usepackage{environ}
\usepackage{latexsym}
\usepackage{amssymb}
\usepackage{epsfig,amsmath,graphics}
\usepackage{epstopdf}
\usepackage{verbatim}
\usepackage{wasysym}
\usepackage{feynmp-auto}
\usepackage{authblk}
\usepackage{xcolor}
\usepackage{enumitem}
\usepackage[utf8]{inputenc}
\usepackage{slashed}
\usepackage{cite}
\usepackage[toc,page]{appendix}
\usepackage[normalem]{ulem}

\usepackage{empheq}

    \newlength\fsep
    \newsavebox\widebox

\definecolor{deepgreen}{rgb}{0.2,0.8,0.3}

\definecolor{deepblue}{rgb}{0.2,0.2,0.8}

\definecolor{deepred}{rgb}{0.8,0.2,0.2}

\usepackage[skip=0pt]{caption}

\begin{document}
\title{The Classical Equations of Motion of Quantized Gauge Theories, Part 2: Electromagnetism}
\author[1]{David E.~Kaplan}
\author[2]{Tom Melia}
\author[1]{Surjeet Rajendran}
\affil[1]{\small Department of Physics \& Astronomy, The Johns Hopkins University, Baltimore, MD  21218, USA}
\affil[2]{\small Kavli IPMU (WPI), UTIAS, The University of Tokyo, Kashiwa, Chiba 277-8583, Japan}

\date{\today}

\maketitle

\begin{abstract}
In this and companion papers, we show that quantum field theories with gauge symmetries permit a broader class of classical dynamics than typically assumed.  In this article, we show that the quantization of electromagnetism  permits the existence of classical electric field states that do not obey Gauss's law. These states are gauge invariant and their time evolution can be consistently described using the Schr\"{o}dinger equation. The time evolution of these states is such that at the classical level, the full set of Maxwell's equations would appear to hold, with the physical effects of these states being attributable to an auxiliary, static ``shadow'' charge density with no internal degrees of freedom. This density could affect the dynamics of charged particles in our universe and it may thus be of observational interest. 
\end{abstract}

\tableofcontents


\section{Introduction}

Quantum electrodynamics, with its better than part-per-billion agreement with experiment, is an extraordinarily successful theory. The central claim of this paper is that despite this success, the conventional quantization procedure used to describe the quantum mechanics of electromagnetism has overlooked a theoretically consistent possibility that likely has interesting observational and theoretical implications. This possibility is the existence of quantum states that violate Gauss's law, but can nevertheless be consistently time evolved using the Schr\"{o}dinger equation in a gauge-invariant manner. The existence of these states is tied to quantum mechanics. In classical electrodynamics, since Gauss's law is one of Maxwell's equations, such a state cannot be described.  However, unlike the classical equations of Maxwell, time evolution in quantum mechanics is described by a single equation, namely, the Schr\"{o}dinger equation. 

When a quantum state evolves, the expectation values of various quantum mechanical operators in that state will automatically obey corresponding classical equations. Since the Schr\"{o}dinger equation is a dynamical equation, it can force the expectation values of quantum states to obey the dynamical equations of the corresponding classical theory, {\it i.e.}, Ampere's law in the case of electromagnetism. But Gauss's law is not a dynamical equation - it is a constraint on the initial states of the theory and is thus not a consequence of the Schr\"{o}dinger equation. The initial state of a quantum theory can be any state that can be described and time evolved in a gauge invariant manner. The main point of this paper is to show that there are such initial states that violate Gauss's law but still lead to gauge invariant physics. We summarize these arguments in Section \ref{sec:summary}. 

In a companion paper \cite{Kaplan:2023wyw}, we pointed out that similar states also exist in General Relativity (also see \cite{Burns:2022fzs}). In that paper, since the quantum theory of General Relativity has been poorly understood \cite{Dirac,PhysRev.160.1113}, we described this point in detail.  In this paper, our interest is electromagnetism - a theory whose quantization is well understood (as it appears in standard textbooks, {\it e.g.}, \cite{Jackiw,Fradkin,Weinberg}). We will thus focus on  the specific arguments in the quantization of electromagnetism to pinpoint how states that violate Gauss's law can also be consistently described in this theory. The existence of these states is most easily understood in the Weyl gauge - we will thus begin our discussion in this gauge in Section \ref{sec:Weyl}. Following this, we show how such states can also be described in other gauges (such as the Coulomb gauge) in  Section \ref{sec:Coulomb}. We then describe the gravitational and cosmological implications of such states in Section \ref{sec:cosmology} and conclude with a discussion about the phenomenological relevance of these states and potential experimental ways to constrain or discover them. 

Throughout we use Greek letters for four-vector indices, Latin letters for three-vector (spatial) indicies, and overdots for coordinate time derivatives. Coordinates ${\bf x}$ and vectors in bold  refer to spatial coordinates and vectors respectively while coordinates $x$ in regular fonts represent space-time coordinates. Throughout this paper, when we describe Hamiltonians, we will work in the Schr\"{o}dinger picture where the operators are time independent and the states evolve in time.

\section{Executive Summary}
\label{sec:summary}

The physics of electromagnetism is contained in the following Lagrangian:

\begin{equation}
\mathcal{L_{EM}} = -\frac{1}{4}F_{\mu \nu}F^{\mu \nu} +  A_{\mu} J^{\mu} + \mathcal{L}_{J}
\label{eqn:QED}
\end{equation}
which describes the interactions of the electromagnetic potential $A_{\mu}$ with a current $J^{\mu}$.\footnote{We have absorbed the electromagnetic coupling $e_c$ in the definition of the current $J^{\mu}$.} The other interactions of this current  (such as its kinetic terms) are described by $\mathcal{L}_{J}$. When we perform a gauge transformation $A_{\mu} \rightarrow A_{\mu} + \partial_{\mu}\alpha$, the kinetic term of this current transforms covariantly so that the Lagrangian \eqref{eqn:QED} is invariant. At the classical level, one obtains the equations of motion by varying the action: 

\begin{equation}
\mathcal{S_{EM}} = \int d^4 x  \, \mathcal{L_{EM}}
\label{eqn:SQED}
\end{equation}
with respect to the gauge field $A_{\mu}$ and setting the variations $\frac{\partial \mathcal{S_{EM}}}{\partial A_{\mu}}$ to zero. Accordingly, we get the following classical equations for the electric (${\bf E}$) and magnetic (${\bf B}$) fields: 

\begin{equation}
    {\bf \nabla}\cdot {\bf E} =  J^{0}
    \label{eqn:Gauss}
\end{equation}
otherwise known as Gauss's law (obtained by varying $A_0$), and
\begin{equation}
    {\bf \nabla} \times {\bf B} - \frac{d {\bf E}}{d t} = {\bf J}
    \label{eqn:Ampere}
\end{equation}
otherwise known as Ampere's law (obtained by varying $A_i$).

Even at the classical level, observe the following fact. Gauss's law \eqref{eqn:Gauss} is a constraint on the initial values of the electric field ${\bf E}$ whereas Ampere's law \eqref{eqn:Ampere} is a dynamical equation. Suppose we ignore \eqref{eqn:Gauss} and solve \eqref{eqn:Ampere} with some initial values for ${\bf E}$ and ${\bf B}$ and the currents $J^{\mu}$ (whose dynamical evolution is governed by its own equation of motion, such as the Lorentz force law).\footnote{While these are second order equations for $A_{\mu}$, they are first order equations for ${\bf E}$ and ${\bf B}$ - thus given initial conditions for ${\bf E}$ and ${\bf B}$ there is no mathematical obstacle to solving these dynamical equations.}  What does this evolution look like? By taking a divergence of \eqref{eqn:Ampere} and using the fact that the dynamical equations of the current $J^{\mu}$ will enforce current conservation $\partial_{\mu}J^{\mu} = 0$, we see that the following identity holds: 

\begin{equation}
    \frac{d \left( {\bf \nabla}\cdot {\bf E} - J^{0}\right) }{d t} = 0
    \label{eqn:preserve}
\end{equation}
Thus, if the initial value of ${\bf E}$ were such that it obeyed Gauss's law $( {\bf \nabla}\cdot {\bf E} - J^{0}) = 0$, the time evolution generated by \eqref{eqn:Ampere} will automatically continue to enforce it. But, suppose we took an initial state of ${\bf E}$ so that Gauss's law was violated by a function of only space, $( {\bf \nabla}\cdot {\bf E} - J^{0}) = J_{s}^0 \neq 0$. Then, the evolution would be such that:
\begin{equation}
    \frac{d \left( {\bf \nabla}\cdot {\bf E} - J^{0}\right) }{d t} = \frac{d J_{s}^{0} }{d t}= 0
    \label{eqn:darkcharge}
\end{equation}
That is, the state looks like the time evolution of a system where in addition to the known current $J^{\mu}$, there is an additional  ``shadow'' charge density $J_{s}^0$ that is somehow unmovable. The interesting fact about a state that violates Gauss's law is that even at the classical level it is described in terms of the gauge invariant observable ${\bf E}$ - there is thus no logical issue involved in time evolving such a state.

At the classical level, we would reject such states simply because we believe in Gauss's law and would thus require $( {\bf \nabla}\cdot {\bf E} - J^{0}) = 0$. But, classical mechanics is not the correct description of nature - the underlying theory is quantum mechanics and classical physics is a limit of quantum mechanics. The key question that we need to ask is if Gauss's law follows from quantum mechanics. We argue that it does not. Instead, we show that quantum mechanics allows for the existence of gauge invariant states of electromagnetism that violate Gauss's law. At the classical level, the time evolution of these states would be identical to that of states containing an unmovable shadow charge $J_{s}^0$ described in \eqref{eqn:darkcharge} - but there is no new degrees of freedom associated with $J_s^0$. It is simply a state of electromagnetism. Further, these states can also be consistently coupled to gravity. 

How could Gauss's law not be true in quantum mechanics but appear to be true in classical electromagnetism? As a first peek at this issue, observe the following. In classical physics, we obtained Maxwell's equations by varying the action $\mathcal{S}_{EM}$ along four independent variations of the potential $A_{\mu}$. But, due to  gauge redundancy, there are not four independent variations of $A_{\mu}$. By Helmholtz's theorem, any potential $A_{\mu}$ can be decomposed as $A_{\mu} = K_{\mu} + \partial_{\mu}\alpha$ where $K_{\mu}$ is  divergence-less, $\partial^{\mu}K_{\mu} = 0$, and thus only contains three degrees of freedom. Write the classical action $\mathcal{S}_{EM}$ in terms of $K_{\mu}$ and $\alpha$ instead of $A_{\mu}$. Due to the gauge invariance of the action $\mathcal{S}_{EM}$ under the gauge transformations $A_{\mu} \rightarrow A_{\mu} + \partial_{\mu}\alpha$ and the associated covariant transformations on  $\mathcal{L}_{J}$, the action: 
\begin{equation}
    \mathcal{S_{EM}} = \int d^4x \, \mathcal{L_{EM}}\left(A_{\mu}, \partial_{\nu} A_{\mu}\right) = \int d^4x \, \mathcal{L_{EM}}\left(K_{\mu}, \partial_{\nu} K_{\mu}\right)
    \label{eqn:lorenzgauge}
\end{equation}
 is only a function of $K_{\mu}$. But since the divergence-less four vector $K_{\mu}$ only has three degrees of freedom, we do not have four independent variations to obtain four independent equations. 

The  reader will observe that in writing \eqref{eqn:lorenzgauge}, we have effectively picked the Lorentz gauge where we set $\partial^{\mu}A_{\mu} = 0$ and as a consequence, we naively lost an equation of motion. The definition of the quantum theory, either at the level of the Hamiltonian or the path integral, requires us to pick a gauge in order to define operators and states. It is this picking of operators and states which affords additional freedom in the derived classical equations. This results in loosening restrictions on the allowed quantum states of the theory, as we'll see, permitting states analogous to \eqref{eqn:darkcharge} that violate Gauss's law. 

\section{Weyl Gauge}
\label{sec:Weyl}

To quantize electromagnetism in Weyl (or temporal) gauge, the following procedure is adopted  \cite{Fradkin} to specify the operators, Hamiltonian and physical states. First, we set $A_0 = 0$ in the classical Lagrangian. The spatial components  $A_j$ of the vector potential and its conjugate momentum  $\Pi_j = \frac{\partial\mathcal{L_{EM}}}{\partial A_j} = -E_j$ get promoted to operators $\hat{A}_j, \hat{\Pi}_j$ with canonical equal time commutation relations: 
\begin{equation}
[\hat{A}_j\left(\bf{x}\right), \hat{\Pi}_{j'}\left(\bf{x'}\right)] = i\, \delta\left({\bf x - x'}\right) \delta_{jj'}
\end{equation}
Using $\Pi_j = -E_j$, we have
\begin{equation}
[\hat{A_j}\left(\bf{x}\right), \hat{E_{j'}}\left(\bf{x'}\right)] = -i\, \delta\left({\bf x - x'}\right) \delta_{jj'}
\end{equation}
The Hamiltonian constructed from these operators is: 
\begin{equation}
    \hat{H}_{W} = \int d^3 {\bf x} \, \left( \frac{1}{2}\left(\hat{\bf {E}}\cdot\hat{\bf {E}} + \hat{\bf {B}}\cdot\hat{\bf {B}}\right) + \hat{\bf {J}}\cdot\hat{\bf {A}}   + \hat{\cal H}_{J}\right)
    \label{eqn:WeylH}
\end{equation}
where $\hat{\bf {B}}\equiv {\bf\nabla}\times\hat{\bf {A}}$ and $\hat{\cal H}_J$ is the remaining Hamiltonian density of the degrees of freedom in the current ${\bf J}$. If a quantum state $|\Psi\rangle$ obeys the Schr\"{o}dinger equation
\begin{equation}
    i \frac{\partial |\Psi\rangle}{\partial t} = \hat{H}_W |\Psi\rangle
    \label{eqn:weylschrod}
\end{equation}
what equations of motion are automatically obeyed by the expectation values of various operators such as $\langle \Psi| {\bf E}| \Psi \rangle$? In the next two subsections, using the canonical and path integral formulations, we will show that Ampere's law follows from the quantum dynamics while Gauss's law does not. Following this discussion, we will show how Gauss's law is obtained in the Weyl gauge and argue that this prescription permits a broader class of quantum states than previously considered.

\subsection{Canonical Formulation}

In the canonical language (see, for example, \cite{Jackiw}), the classical equations should be true in expectation value (Ehrenfest's theorem).  This can be seen by taking a time derivative of expectation values of fields and using the Schr\"{o}dinger equation:
\begin{eqnarray}
    \partial_t\langle\hat{E}^j({\bf x})\rangle &=& i \langle \left[\hat{H}_W,\hat{E}^j({\bf x})\right]\rangle \nonumber\\
    &=& \int d^3 {\bf x}' \, \langle \frac{1}{2} \left[({\bf \nabla}\times\hat{\bf {A}}({\bf x}'))^2 , {\hat{E}}^j({\bf x}) \right] + \left[ \hat{\bf {A}}({\bf x}'),\hat{{E}}^j({\bf x}) \right] \cdot \hat{\bf {J}}({\bf x}') \,\rangle \nonumber\\ \label{eq:Ampere}
    &=&   \langle ({\bf \nabla} \times \hat{\bf {B}} )^j({\bf x})\rangle - {\langle\hat{{J}}^j({\bf x})\,\rangle}
\end{eqnarray}
where we integrated by parts to generate the last step.  We see here that Ampere's law is reproduced in expectation value.  This, and $\partial_t \langle \hat{\bf {A}}\rangle = -\langle \hat{\bf {E}} \rangle$, are the only equations of motion for the electromagnetic field predicted by this method.  

But what about Gauss's law?  The absence of Gauss's law reflects the missing conjugate momentum of $A_0$, and thus the missing equation of motion.  As in the classical case, a divergence of \eqref{eq:Ampere} reproduces the time-derivative of the expectation value of Gauss's law (using $\langle \partial_\mu J^\mu \rangle = 0$):

\begin{equation}
    \partial_t \langle {\bf\nabla}\cdot\hat{\bf E}\rangle = - \langle {\bf\nabla} \cdot \hat{\bf J} \rangle = \partial_t\langle \hat{J}^0 \rangle
\end{equation}
At this point, the case is usually made to impose Gauss's law by fixing the remaining spatial gauge invariance.  Classically, in Weyl gauge, the spatial transformations ${\bf A}({\bf x}) \rightarrow {\bf A}({\bf x}) + {\bf\nabla}\alpha ({\bf x}) $ are still an invariance of the action.  Quantum mechanically, the generator of this transformation is the operator $\hat{G} \equiv {\bf\nabla}\cdot\hat{\bf E} - \hat{J}^0$.  Not surprisingly, it can be checked that this operator commutes with the Hamiltonian:
\begin{equation}
    \left[ \hat{G},\hat{H}_W\right] = 0
\end{equation}
which means this operator and the Hamiltonian can be simultaneously diagonalized.  One can now impose Gauss's law by requiring physical state vectors to be invariant under spatial gauge transformations, namely:
\begin{eqnarray}
    &\hat{G}\,|\Psi_{EM}\rangle = 0 \\ &  \nonumber \\
   & e^{-i\int d^3 {\bf x} \,\alpha\, \hat{G} } \,|\Psi_{EM}\rangle = |\Psi_{EM}\rangle
\end{eqnarray}
and thus all physical states are constrained to obey Gauss's law.  Thus we see, to obtain Gauss's law, a constraint equation, from the quantum field theory, one has to impose it by hand on the physical states.  We will see this again below in the path-integral formulation.  Then we will see that this constraint on states is not a requirement for a consistent quantum theory.


\subsection{Path Integral Formulation}

It is useful to reproduce this in the path integral language (see, for example, \cite{Fradkin}).  Let us {\it  naively} construct the  path integral that solves \eqref{eqn:weylschrod} in the field basis: 
\begin{equation}
      \mathcal{T} = \langle  {\bf A}_f | T\left(t_2; t_1\right) |  {\bf A}_{i}\rangle = \int_{{\bf A}\left(t_1\right) = {\bf A}_i}^{{\bf A}\left(t_2\right) = {\bf A}_f}\, DA_\mu \, D\lambda \,  e^{i \int_{t_1}^{t_2} d^4 x  \, \left(\mathcal{L_{EM}} -\lambda A_0\right)}
    \label{eqn:PIWeyl}
\end{equation}
where the Lagrange multiplier $\lambda$ enforces the Weyl gauge $A_0 = 0$. This path integral yields the transition matrix element for the field basis state $|{\bf A}_{i}\rangle$ at time $t_1$ to evolve to $| {\bf A}_f\rangle$ at time $t_2$. This transition matrix element \eqref{eqn:PIWeyl} should be invariant when we compute this path integral with a variable redefinition $A_{\mu} \rightarrow A_{\mu} + \delta A_{\mu}$ with $\delta A_{\mu}$ vanishing at the boundaries. This yields a set of Schwinger-Dyson equations which show how the classical field equations arise as identities automatically obeyed by the expectation values of various field operators when the quantum state evolves as per \eqref{eqn:PIWeyl}. One can check that when this procedure is applied to the spatial variations $\delta A_i$, one obtains the result that Ampere's law is obeyed by the expectation values of the quantum operators. But, for the variation $\delta A_0$, this yields the equation: 
\begin{equation}
    \langle \Psi | {\bf\nabla}\cdot{\bf E} - J^0 + \lambda |\Psi\rangle = 0
\end{equation}
 This is not an equation of motion or a constraint on the physical state $|\Psi\rangle$ - instead, it describes how the unphysical Lagrange multiplier $\lambda$ evolves in the path integral to maintain the gauge $A_0 = 0$. Thus Gauss's law does not immediately follow from the quantum Hamiltonian  $\hat{H}_{W}$ .

 \subsection{Gauss's Law and Its Violation}
  The origin of Gauss's law in the Weyl gauge is tied to the elimination of residual spatial gauge transformations in the theory. The gauge choice $A_0 = 0$ does not eliminate all the gauge freedom in the theory - in principle we would like to identify quantum states  $|{\bf A}\rangle$ that are related to each other by purely spatial gauge transformations: $|{\bf A}\rangle \equiv |{\bf A} + {\bf\nabla}\alpha\rangle$ for any $\alpha\left({\bf x}\right)$. For this equivalence to hold, we need the following to be true:
 \begin{equation}
 \langle  {\bf A}_f | T\left(t_2; t_1\right) |  {\bf A}_{i}\rangle = \langle  {\bf A}_f + {\bf\nabla}\alpha_f | T\left(t_2; t_1\right) |  {\bf A}_{i} + {\bf\nabla}\alpha_i \rangle
 \label{eqn:strongGauge}
 \end{equation}
 with $\alpha_i \left({\bf x}\right) \neq \alpha_f\left({\bf x}\right)$. It can be checked that when $\alpha_i \left({\bf x}\right) \neq \alpha_f\left({\bf x}\right)$, the path integral  \eqref{eqn:PIWeyl} does not maintain this equality, violating the equivalence $|{\bf A}\rangle \equiv |{\bf A} + {\bf\nabla}\alpha\rangle$.  To maintain this equivalence, the physical Hilbert space of the theory is restricted to a smaller space of states that are  invariant under these spatial gauge transformations and the time evolution operator \eqref{eqn:PIWeyl} is projected onto this restricted space, leading to physics that is also invariant under such general spatial gauge transformations.

 Let us see how this works. We want the physics of the states $|{\bf A}\rangle$  to be identical to that of the states  $|{\bf A} + {\bf\nabla}\alpha\rangle$. As we've seen, these spatial gauge transformations are generated by the operator $\hat{G}={\bf\nabla}\cdot\hat{{\bf E}} - \hat{J}^0$. 
 Thus, under a spatial gauge transformation, the eigenstates of this operator will transform with an overall,  physically irrelevant,  phase. Now, consider the physics of a subspace of the Hilbert space where eigenstates of $\hat{G}$ with the same eigenvalue. Since the operator $\hat{G}$ commutes with $\hat{H}_W$, the time evolution of an initial state in this subspace will remain in the same subspace. Together, these facts imply that the physics of this subspace is invariant under spatial gauge transformations. 
 
 In the traditional quantization procedure ({\it e.g.}, \cite{Fradkin}), the physical states $|\Psi_{EM}\rangle$  are taken to be eigenstates of  $\hat{G}$ with zero eigenvalue. In this subspace, Gauss's law is preserved. But, this is a choice. In a subspace of eigenstates of $\hat{G}$ with a non-zero eigenvalue (function) $ J_{s}^{0}\left({\bf x}\right)$,
\begin{equation}
    \hat{G}({\bf x})|\Psi_{EM}\rangle = J_{s}^0\left({\bf x}\right) |\Psi_{EM}\rangle
    \label{eqn:WeylViolateGauss}
\end{equation}
the states all transform with the same phase:
\begin{equation}
    e^{-i\int d^3 {\bf x} \,\alpha\, \hat{G} } \,|\Psi_{EM}\rangle = e^{-i\int d^3 {\bf x} \,\alpha\, J_s^0 }|\Psi_{EM}\rangle
\end{equation}
which thus also leads to gauge invariant physics.  For these states, Gauss's law is not obeyed. Instead, the time evolution is such that
\begin{equation}
    \frac{d\langle \Psi_{EM} | \,\hat{G}\,|\Psi_{EM}\rangle } {dt}  = 0
\end{equation}
which is exactly the form of \eqref{eqn:darkcharge}. 
 
 We thus see that there are gauge invariant quantum states that violate Gauss's law and there are no difficulties in time evolving these states. The initial quantum state of the universe could have been a state where Gauss's law was preserved, {\it i.e.}, a state that was an eigenstate of $\hat{G}$ with eigenvalue zero. But it could just as easily have been an eigenstate of  $\hat{G}$ with a non-zero eigenvalue $J_{s}^0 \left( {\bf x}\right)$. In this case, Gauss's law would be violated, and it is a matter for experiment to decide which of these scenarios is realized in our universe.  We are simply choosing a Hilbert space in which the states have a background, static, longitudinal electric field.  While such a field chooses a rest frame (and generally breaks Poincare invariance), the current state of dynamical degrees of freedom ({\it e.g.}, photons and electrons) also seem to choose a rest frame.  The only difference is that we have turned on something additional that evolves trivially.

 It is possible that it is not necessary to fix the remaining spatial gauge symmetry.  It would remain a time-independent symmetry of the Hamiltonian and thus should not generate, for example, a photon mass. Without requiring this fix, it should be possible to write physical states which are not eigenstates of $\hat{G}$.  While it is not clear this is important physically, it may play an important role in the non-Abelian version of our story, which we will explore in the next paper.


\section{Other Gauges}
\label{sec:Coulomb}
In the Weyl gauge, we have shown the existence of gauge invariant quantum states that violate Gauss's law. In this section, we show how they can be described in other gauges, such as the Coulomb gauge. 

To describe this construction, we begin by reviewing the conventional methods used to translate the physics of electromagnetism in the Weyl gauge to other gauge choices.  Since this is a review, we simply sketch the major steps and refer the reader to \cite{Fradkin} for the details. The first step in this procedure is to construct the Lagrangian that describes the physics of the Weyl gauge from the Hamiltonian  $\hat{H}_W$ in \eqref{eqn:WeylH}. As described in Section \ref{sec:Weyl}, we restrict the Hilbert space - to states that are eigenstates of  $\hat{G} = {\bf \nabla}\cdot\hat{{\bf E}} - \hat{J}^0$ with the same eigenvalue making the theory trivial under spatial gauge transformations. We can construct the path integral \eqref{eqn:PIWeyl} while restricting to such gauge-invariant states. In the conventional quantization procedure, this is enforced by inserting a projection operator $\hat{P}$ that projects the basis states $|\hat{{\bf A}}\rangle$ onto the eigenspace of zero eigenvalue of $\hat{G}$. Thus, the generating functional of the theory is: 

\begin{equation}
\mathcal{Z} = \text{tr}\left( T e^{-i \int dt \hat{H}_{W}} \hat{P}\right)
\end{equation}
where the projection operator 
\begin{equation}
\hat{P} = \Pi_{t,{\bf x}} \,  \delta \left(\hat{G} \left( {\bf x}\right) \right) 
\label{eqn:delta}
\end{equation}
This projection operator acts on each point in time and it can be implemented in the path integral using an integral representation of the delta function  \eqref{eqn:delta}

\begin{equation}
\delta \left({\bf \nabla}\cdot{\bf E}\left( {\bf x}, t\right) - {J}^0 \left( {\bf x}, t\right) \right)  = \int DA_{0} \,  e^{i \delta t \int d^3 {\bf x} \, A_0 \left( {\bf x}, t\right) \left({\bf \nabla}\cdot {\bf E}\left({\bf x}, t\right) - J^{0}\left( {\bf x}, t\right) \right) }
\end{equation}
We thus see the role of $A_0$ - it enforces the constraint that the Hilbert space of the theory is restricted to a specific space of states that are annihilated by the operator $\hat{G}$. Constructing a path integral via the standard procedure of inserting complete sets of states, and including the above delta function, one obtains the action: 
\begin{equation}
S\left(A_{\mu}, E\right) = \int d^4 x \left( -{\bf E} \cdot \partial_t {\bf A}  - \frac{1}{2}\left(  {\bf E}^2 + {\bf B}^2\right) - {\bf A}\cdot {\bf J}  + A_0 \left( {\bf \nabla}\cdot {\bf E} - J^0\right) + \cdots \right)
\end{equation}
where the ellipsis represents the remaining matter terms.  From this, one performs the Gaussian path integral over $E$ and other conjugate variables and obtains the conventional Lagrangian \eqref{eqn:QED}. 

Let us see how this procedure would change if we had picked a subspace of states where $\hat{G}$ has a non-zero eigenvalue $J_{s}^0\left( {\bf x}\right)$. In this case, the projection operator would be: 

\begin{equation}
\hat{P} = \Pi_{t,{\bf x}} \,  \delta \left(\hat{G}\left( {\bf x}\right) - J_{s}^{0}\left( {\bf x}\right) \right) 
\label{eqn:ourdelta}
\end{equation}
The subsequent mathematical procedure  (following the steps in \cite{Fradkin}) results in the effective Lagrangian: 

\begin{equation}
\mathcal{\tilde{L}_{EM}} = -\frac{1}{4}F_{\mu \nu}F^{\mu \nu} +  A_{\mu} J^{\mu} + A_{\mu}J_{s}^{\mu}+ \mathcal{L}_{J}
\label{eqn:ourQED}
\end{equation}
where the field $J_{s}^{\mu}$ is the background classical field $J_{s}^{\mu}=\left(J_{s}^0\left( {\bf x}\right), 0, 0, 0\right)$. Thus, the theory is identical to that of electromagnetism coupled to a classical background charge density $J_{s}^0\left( {\bf x}\right)$ - this charge density picks a rest frame and it is thus the theory of electromagnetism in such a Lorentz breaking background. While it is well recognized that such a classical background can be added to quantum electrodynamics, the key point of our paper is to point out that there is no additional microphysics associated with this background. The field theory that describes $J_{s}^0\left( {\bf x}\right)$ is simply a quantum state of electromagnetism. 

The quantum theory of states with  $J_{s}^0\left( {\bf x}\right) \neq 0$ in any gauge can now be described by applying the conventional quantization procedure specific to that gauge starting with the effective Lagrangian \eqref{eqn:ourQED}.  For example, to obtain the quantum theory in Coulomb gauge, construct the canonical Hamiltonian corresponding to the effective Lagrangian \eqref{eqn:ourQED}. Then, impose the operator requirement  ${\bf \nabla}\cdot\hat{{\bf A}} = 0$ and solve for  $\hat{A}_0$  in terms of the effective charge density  $\hat{J}^0 + J_{s}^0$: 

\begin{equation}
    \hat{A}_0\left( {\bf x}\right) = \int d^3 {\bf x'} \frac{\left(\hat{J}^{0}\left({\bf x'}\right) + J_{s}^0\left({\bf x'}\right)\right)}{4 \pi {\bf | x - x'|}}
    \label{eqn:coulomb}
\end{equation}
The resulting Hamiltonian describes the quantum theory of states with  $J_{s}^0\left( {\bf x}\right) \neq 0$.


\section{Gravitation and Cosmology}\label{sec:cosmology}


In this section, we  describe how states that violate Gauss's law couple to gravity. Our treatment of the gravitational interactions parallels the flat space treatment - we begin by describing these states in the Weyl gauge where the existence of these states is most easily understood. From the Weyl gauge, we construct the effective Lagrangian that describes these states and the gravitational dynamics can be readily read off from this Lagrangian. 

Our principal interest here is to understand the cosmological implications of such states. We will thus specialize to the case of a FRW cosmology - but, the methods we describe can be extended to any space-time. Accordingly, we take the metric of the space-time from the interval: 

\begin{equation}
     ds^2 =    - N(t)^2dt^2 + a(t)^2\left(dx^2 + dy^2 + dz^2\right)
\end{equation}
where  $a(t)$ is the scale factor. $N(t)$ is a gravitational gauge degree of freedom that needs to be fixed in order to define a Hamiltonian \cite{Kaplan:2023wyw, Burns:2022fzs}. While we will eventually fix $N(t)$, we retain it for now for the sake of clarity.  Since we know how regular charged particles behave in the presence of these states, we will focus our attention solely on the cosmological evolution of these states, neglecting the physics of charged matter that may be coupled to these states. The Lagrangian that describes this system is: 

\begin{equation}
 \mathcal{L_{T}} =  \sqrt{-g} \left( -\frac{1}{4} F_{\mu \nu}F^{\mu \nu}+  \frac{M_{pl}^2}{2}R \right) 
\end{equation}

With this Lagrangian,  we obtain the conjugate momenta: 

\begin{equation}
\Pi^j = \frac{\partial \mathcal{L_{T}}}{\partial \dot{A_j}}
\end{equation}
for the electromagnetic degrees of freedom and 

\begin{equation}
\Pi_a = \frac{\partial \mathcal{L_{T}}}{\partial \dot{a}}
\end{equation}
for the scale factor. We pick the Weyl gauge by setting  $A_0 = 0$. We then set equal time  commutation relations: 

\begin{equation}
[\hat{a}, \hat{\Pi}_{a}] = i 
\nonumber
\end{equation}

\begin{equation}
[\hat{A}_j\left({\bf x}\right), \hat{\Pi}^{j'}\left({\bf x'}\right)] = i \delta_{j}^{j'} \delta\left( {\bf x - x'}\right)
\label{eqn:cosmiccommutator}
\end{equation}
With these definitions, the electric field operator is: 

\begin{equation}
\hat{E}_i = \hat{F_{0i}} = N\left(t\right) \frac{\hat{\Pi}_i}{\hat{a}}
\end{equation}
and the Hamiltonian for the full system (gravity and electromagnetism) is\footnote{A discussion of the subtleties of defining operators with inverse scale factors are discussed in \cite{Kaplan:2023wyw}}: 

\begin{equation}
\hat{H}_{T} = N\left(t\right) \left(\int d^3 {\bf x}  \left( \frac{ \hat{\bf{\Pi}}^2 +  \hat{\bf{B}}^2 }{2 \, \hat{a}}\right)  - \frac{1}{12 \, M_{pl}^2} \frac{\hat{\Pi}_a^2}{ \hat{a}}\right)
\label{eqn:FullH}
\end{equation}
where we define ${\bf B}$ using the Maxwell tensor $F_{\mu \nu}$. The time evolution of physical states $|\Psi\rangle$ is given by the equation: 

\begin{equation}
i\,\frac{\partial |\Psi\rangle}{\partial t} = \hat{H}_{T} |\Psi\rangle
\end{equation}

Similar to the situation in flat space, notice that ${\bf\nabla}\cdot{\hat{\bf\Pi}}$ commutes with the Hamiltonian $\hat{H}_{T}$. Further, due to  the commutation relations \eqref{eqn:cosmiccommutator},  ${\bf\nabla}\cdot\hat{\bf\Pi}$ is  also the generator of  spatial gauge transformations, where ${\bf\nabla}$ is the gradient operator on the comoving coordinates ${\bf x}$. Following the arguments in Section \ref{sec:Weyl}, to obtain gauge invariant physics, we demand that the physical states $|\Psi\rangle$ are eigenstates of the operator ${\bf\nabla}\cdot\hat{\bf{\Pi}}$, {\it i.e.}: 

\begin{equation}
{\bf{\nabla}}\cdot\hat{\bf{\Pi}} |\Psi\rangle = J_{s}^0 \left( {\bf x}\right) |\Psi\rangle
\end{equation}
The above equations fully describe how these states interact with gravitation for any choice of time parameterization $N(t)$.  For simplicity, we will choose $N(t)=1$.

Now we can examine what the states that violate Gauss's law look like in an expanding background. To understand this physics, we want to  parallel the discussion in Section \ref{sec:Coulomb} and obtain the effective Lagrangian that describes these states in a covariant manner. For simplicity, we assume that the energy density in these states is small and that the dynamical evolution of the scale factor $a(t)$ is governed by other, more dominant energy densities. We are thus interested in understanding how these states respond to a pre-determined dynamical evolution of $a(t)$, {\it i.e.}, we want to understand how these states ``redshift'' in the semi-classical limit. With these assumptions, the physical state $|\Psi\rangle = |a_{cl}\rangle\otimes|\Psi_{EM}\rangle$ where $|a_{cl}\rangle$ describes the state of the metric and $|\Psi_{EM}\rangle$ is the state of the electromagnetic fields. $|a_{cl}\rangle$ is a coherent state of the gravitational field and in the semi-classical limit that we are in, we can simply replace the operators $\hat{a}$ and $\hat{\Pi}_a$ in \eqref{eqn:FullH} with their corresponding classical field values $a(t)$ and $\Pi_{a}(t)$. 

We can now easily adapt our discussion to the discussion in Section \ref{sec:Coulomb}. Accordingly, for the generating functional: 

\begin{equation}
\mathcal{Z} = \text{tr}\left( T e^{-i \int dt \hat{H}_{T}} \hat{P}\right)
\end{equation}
with the projection operator: 
\begin{equation}
\hat{P} = \Pi_{t,{\bf x}} \,  \delta \left({\bf \nabla}\cdot\hat{\bf {\Pi}}( {\bf x}) - J_{s}^0 ( {\bf x}) \right) 
\end{equation}
Enforcing this projection operator via the integral representation of the delta function and proceeding as in Section \ref{sec:Coulomb}, we get the effective Lagrangian: 

\begin{equation}
\mathcal{\tilde{L}_{EM}} = \sqrt{-g}\left( -\frac{1}{4} F_{\mu \nu}F^{\mu \nu} + A_{\mu}{\mathbb J}^{\mu}\right)
\label{eqn:GRGauss}
\end{equation}
where the current ${\mathbb J}^{\mu}$ is of the form: 
\begin{equation}
{\mathbb J}^{\mu} = \left(\frac{J_{s}^0\left({\bf x}\right)}{a(t)^3}, 0, 0, 0\right)
\end{equation}
Notice that the current density ${\mathbb J}^{\mu}({\bf x})$ obeys the conservation equation: 

\begin{equation}
\nabla_{\mu}{\mathbb J}^{\mu} = 0 
\label{eqn:conservedGRcurrent}
\end{equation}
where $\nabla_\mu$ is the covariant derivative with respect to the background metric.

 The physical effects of these states  can be computed using the effective Lagrangian \eqref{eqn:GRGauss} in concert with the conservation equation \eqref{eqn:conservedGRcurrent} - these effects are identical to that of a conserved, static background charge density that redshifts with the expansion of the universe. For example, breaking $\hat{\bf \Pi}$ into transverse and longitudinal parts, $\hat{\bf \Pi} = \hat{\bf \Pi}_\perp + {\bf\nabla}\hat{\Phi}$, the expectation value of the Hamiltonian contains a contribution from the longitudinal part:
 \begin{equation}
     \langle\Psi | \hat{H}_T |\Psi\rangle \supset - \frac{1}{2 a(t)}\int d^3{\bf x}\, d^3 {\bf x'}\, \frac{J_s^0({\bf x'}) J_s^0({\bf x})}{4\pi |{\bf x} - {\bf x'}|} \equiv  \frac{V_{coulomb}^s}{a(t)}
 \end{equation}
 and we see that the effective Coulomb potential energy of the shadow charge density redshifts as $a^{-1}$, as it would for any two charges a fixed coordinate distance apart.
 
 One can also read off the redshift from the energy-momentum tensor.  In the Weyl gauge, the time-time component of the energy-momentum tensor is:
 \begin{eqnarray*}
     T_0^0 &=& F_{0\mu} F^{\mu 0} - \frac{1}{4}\delta_0^0 F_{\alpha\beta} F^{\alpha\beta} \\
        &=& - \frac{1}{2}\frac{F_{0i}^2}{a(t)^2} = - \frac{{\bf \Pi}^2}{2 a(t)^4}
 \end{eqnarray*}
 where in the second line we used the quantum relation between $\hat{\Pi^i}$ and $\hat{F}_{0i}$ and inserted the classical expectation values.  For the Gauss's law violation ${\bf\nabla}\cdot{\bf\Pi} = J_s^0$, the $\bf\Pi$ is time-independent and thus the energy density redshifts as radiation.

\section{Discussion}

In this paper, we have shown that the theory of quantum electrodynamics permits a broader class of  quantum states that can be time evolved in a gauge invariant manner than traditionally considered. These states violate Gauss's law, leading to electric fields in the universe that act as though they were sourced by an immovable, conserved background classical charge density. But there is no additional micro-physics associated with this charge density - these are simply allowed states of electromagnetism. These states pick a rest frame and thus break Lorentz invariance.  

Since there is nothing logically wrong with such states, it is a matter of experiment and observation to see if our universe is in a quantum state where Gauss's law is preserved or violated. It is likely that there would be rich phenomenology associated with these states since the effective ``shadow'' charge density that is associated with these states does not represent new dynamics or degrees of freedom. The latter is typically constrained by a variety of stringent astrophysical and collider limits, limiting its ability to significantly impact standard model particles. But, these are simply states of electromagnetism and thus they can have a significant impact on standard model particles without being subject to such constraints. It would be interesting to develop the phenomenology of these states and identify their cosmological, astrophysical and laboratory signatures. This is especially important since the discovery of such states would rule out a period of cosmic inflation in the past history of our universe. 

For simplicity, in this paper, we also assumed that the rest frame picked by these states was the same as the cosmic rest frame. It would be interesting to see if novel phenomenology could arise if this is not the case. It might be that the relative motion between the cosmological background and these states would effectively source long range, coherent magnetic fields and these might be of cosmological significance. In addition to cosmology, it would also be interesting to develop the phenomenology of these states in the vicinity of black hole horizons. Since these states act as a fixed background, it is plausible that they source new physical divergences around such horizons. This may provide new opportunities to discover such states and potentially lead to additional insight into divergences caused by quantum mechanics \cite{Kaplan:2018dqx,Hollands:2019whz}  in the vicinity of horizons. 

While we have focused on the existence of these broader class of quantum states in electromagnetism (in this paper) and gravitation (in our companion paper \cite{Kaplan:2023wyw}), it is likely that such states also exist for non-Abelian gauge theories. These theories also possess non-dynamical degrees of freedom and their quantum description requires suitably fixing various gauge degrees of freedom. It is thus possible these theories also allow for the existence of such ``shadow'' non-Abelian charge densities in a cosmological rest frame. In future work, we intend to develop the theoretical framework to analyze these effects and extract the associated phenomenology.

\section*{Acknowledgements}
We thank Michael Peskin and  Raman Sundrum for fruitful discussions. This work was supported by the U.S.~Department of Energy~(DOE), Office of Science, National Quantum Information Science Research Centers, Superconducting Quantum Materials and Systems Center~(SQMS) under Contract No.~DE-AC02-07CH11359. D.E.K.\ and S.R.\ are supported in part by the U.S.~National Science Foundation~(NSF) under Grant No.~PHY-1818899.
S.R.\ is also supported by the~DOE under a QuantISED grant for MAGIS.
The work of S.R.\  was also supported by the Simons Investigator Award No.~827042. T.M.\ is supported by the World Premier International Research Center Initiative (WPI) MEXT, Japan, and by JSPS KAKENHI grants JP19H05810, JP20H01896,  JP20H00153, and JP22K18712. 
\bibliographystyle{unsrt}
\bibliography{referencesEM}

\end{document}